\documentclass[12pt]{article}
\usepackage{epsfig}

\newcommand{\lsim}{\raisebox{-4pt}{$\,\stackrel{\textstyle
                                                         <}{\sim}\,$}}

\newcommand{\nn}{\nonumber}
\newcommand{\gev}{{\rm GeV}}
\def\mev{\,{\rm MeV}}

\newcommand{\als}{\alpha_{\rm s}}
\newcommand{\aem}{\alpha_{\rm em}}
\newcommand{\vk}{{\bf k}_\perp}
\newcommand{\vd}{{\bf \Delta}_\perp}

\newcommand{\be}{\begin{equation}}
\newcommand{\ee}{\end{equation}}
\newcommand{\ba}{\begin{eqnarray}}
\newcommand{\ea}{\end{eqnarray}}

\begin{document}
\thispagestyle{empty}
\begin{flushright}
WU B 01-07 \\
hep-ph/yymmxxx\\
October 2001\\[5em]
\end{flushright}

\begin{center}
{\Large \bf Perturbative and Non-Perturbative QCD Corrections\\[0.5em] to\\[0.5em] 
                       Wide-Angle Compton Scattering}\\

\vskip 3\baselineskip

H.W.\ Huang$^a$\footnote{Postdoctoral research fellow (No. P99221) of 
the Japan Society for the Promotion of Science (JSPS).}, 
P. Kroll$^b$ and T. Morii$^a$\\[0.5em]
{\small{\it $a$ Faculty of Human Development, Kobe University, Nada,
Kobe 657-8501, Japan}}\\
{\small{\it {$b$} Fachbereich Physik, Universit\"{a}t Wuppertal,
D-42097 Wuppertal, Germany}}\\
\vskip \baselineskip
\end{center}

\vskip 3\baselineskip
\begin{abstract}
We investigate corrections to the handbag approach for wide-angle
Compton scattering off protons at moderately large momentum transfer:
the photon-parton subprocess is calculated to next-to-leading order in
$\alpha_s$ and contributions from the generalized parton
distribution ${E}$ are taken into account. Photon and proton 
helicity flip amplitudes are non-zero due to these corrections which 
leads to a wealth of polarization phenomena in Compton scattering. 
Thus, for instance, the incoming photon asymmetry or the transverse
polarization of the proton are non-zero although small.
\end{abstract}

\bigskip


\newpage
\section{Introduction} 
\begin{figure}[t]
\begin{center}
\epsfig{figure=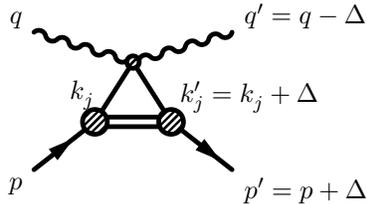, bb= 215 665 370 775, width=5.0cm, clip=}
\end{center}
\vspace{-0.4cm}
\caption{\label{fig1} The handbag diagram for Compton scattering off
  protons. The horizontal lines represent any number of spectator
  partons. }
\end{figure}

Probing the proton with high-energy photons provides information about
its inner structure. The most famous process used for such
investigations is deep inelastic lepton-proton scattering. From a
dynamical point of view this process represents forward (virtual)
Compton scattering and is described by the handbag diagram shown in Fig.\
\ref{fig1}. Recent theoretical developments revealed that the physics
of the handbag diagram is also of importance for deeply virtual
\cite{rad97,ji-osb} and wide-angle \cite{rad98,link} Compton
scattering off protons. Both these processes refer to complementary
kinematical situations. The region of deeply virtual scattering is
characterized by small momentum transfer from the initial to the final 
proton and a large photon virtuality while in the wide-angle region
the situation is reversed. As
has been argued in \cite{rad98,link} the wide-angle Compton amplitudes
approximately factorize into hard photon-parton subprocess amplitudes
and proton matrix elements representing the soft emission and
reabsorption of a parton by the proton. These matrix elements are
moments of generalized parton distributions (GPDs)
\cite{rad97,ji97,mue94} and can be regarded as new form factors of the
protons. The GPDs also encode the soft physics information required to
describe deeply virtual Compton scattering. That the handbag diagram,
i.e.\ elastic scattering of photons from quarks, controls Compton
scattering has been conjectured by Bjorken and Paschos \cite{bjo} and
by Scott \cite{scott} long time ago as we note in passing. 

It is however to be emphasized that the handbag contribution to
wide-angle Compton scattering formally represents only a power
correction to the leading-twist perturbative contribution \cite{bro80}. 
This contribution for which all partons the proton is composed
participate in the hard scattering and not only a single one as in the
handbag, has been calculated several times \cite{farrar} with partially
contradicting results. According to the most recent study \cite{dixon}, 
it seems difficult to account for the wide-angle data on Compton
scattering \cite{shupe}. This result as well as similar observations
made with the pion and the proton electromagnetic form factors 
\cite{pion,bolz} have lead to the assumption of a dominant handbag 
contribution for momentum transfers below about 100 GeV$^2$. 
There is a third contribution to Compton scattering. It has the
topology of the so called cat's ears graphs where the hard subprocess
involves two partons. It is reasonable to assume that the magnitude of
this contribution is intermediate between the handbag and the
perturbative one and that it can be neglected in the kinematical
range of interest. 

In this work we are going to investigate perturbative and
non-perturbative QCD corrections to the handbag contribution for
wide-angle Compton scattering. We calculate the next-to-leading order
(NLO) corrections to the subprocess and, motivated by the surprising
result for the Pauli form factor found at JLab \cite{jon00}, we study  
the bearing of the in \cite{rad98,link} neglected form factor $R_T$ 
on the predictions. We begin with a sketch of the handbag approach and 
the calculation of the NLO corrections (Sect.\ 2). A brief discussion
of the model used for the form factors, or the underlying 
GPDs, follows (Sect.\ 3). Sect.\ 4 is devoted to a comprehensive
discussion of the predictions for a large set of observables and their 
comparison with the results presented in \cite{link} and with those 
obtained  with other theoretical concepts. This may facilitate
the interpretation of future experimental data on wide-angle Compton 
scattering that might be obtained at Spring-8, JLab or at an ELFE-type
accelerator. Finally we discuss the possibility of measuring 
the Compton form factors (Sect.\ 5) and close with a summary (Sect.\ 6).
\section{The handbag contribution}
\label{sect2}
Let us sketch the calculation of the handbag contribution to
wide-angle Compton scattering; for details we refer to \cite{link}. 
For Mandelstam variables, $s$, $t$ and $u$, that are large on a hadronic
scale, $\Lambda^2$, of the order of 1 GeV$^2$, the Compton amplitudes are 
calculated from the handbag graph displayed in Fig.\ \ref{fig1}. Its 
contribution is defined through the assumption that the soft hadron
wave functions occurring in the Fock decomposition of the proton, are 
dominated by parton virtualities in the range $|k_i^2| \lsim \Lambda^2$ 
and by intrinsic transverse parton momenta $\vk{}_i$ that satisfy 
$\vk{}_i^2/x_i \lsim \Lambda^2$. The intrinsic transverse momentum of a 
parton is defined in a frame where its parent's hadron transverse 
momentum is zero; $x_i=k_i^+/p^+$ is the usual light-cone momentum 
fraction. It is of advantage to choose a symmetric frame of reference 
where the plus and minus light-cone components of the momentum
transfer, $\Delta$, are zero (for the definition of the kinematics see 
Fig.\ \ref{fig1}). This implies $t=-\vd$ as well as a vanishing 
skewedness parameter, $\xi=(p-p')^+/(p+p')^+$. One can then show that 
the photon-parton scattering is hard and the momenta
$k_j^{\phantom{.}}$, $k'_j$ of the active partons, i.e.\ those to
which the photons couple (see Fig.\ \ref{fig1}), are approximately 
on-shell, collinear with their parent hadrons and with momentum
fractions $x_j^{\phantom{.}} = x'_j = 1$. This leads to an approximate 
equality of the Mandelstam variables in the photon-parton subprocess 
and in the overall photon-proton reaction up to corrections of order 
$\Lambda^2/t$. 

In view of this the helicity amplitudes ${\cal M}_{\mu'\nu',\,\mu \nu}$ 
of wide-angle Compton scattering in the symmetric frame are given by 
\ba
{\cal M}_{\mu'+,\,\mu +}(s,t) &=& \;2\pi\aem 
            \left[{\cal H}_{\mu'+,\,\mu+}(s,t)\,(R_V(t) + R_A(t))\,
  + \, {\cal H}_{\mu'-,\,\mu-}(s,t)\,(R_V(t) - R_A(t)) \right ]\,, \nn\\[0.5em]
{\cal M}_{\mu'-,\,\mu +}(s,t) &=& \;-\pi\aem \frac{\sqrt{-t}}{m} 
         \left[{\cal H}_{\mu'+,\,\mu+}(s,t)\, 
         + \, {\cal H}_{\mu'-,\,\mu-}(s,t)\, \right] \,R_T(t)\,.
\label{final}
\ea
Here, $\mu$ ($\nu$) and $\mu'$ ($\nu'$) denote the light-cone
helicities \cite{kogut,diehl01} of the incoming and outgoing
photon (proton), respectively. $m$ is the proton mass. For the sake of
legibility explicit helicities are labeled only by their signs. We
emphasize that the proton helicity flip amplitudes have been neglected
in Refs.\ \cite{rad98,link}. Below we will discuss under which
circumstances this is reasonable and when not.

The soft proton matrix elements, $R_i$ ($i=V,A,T$), appearing in 
Eq.\ (\ref{final}) represent new types of proton form factors. They
are defined as $1/x$-moments of GPDs at zero skewedness. For active
quarks of flavour $a$ ($u$, $d$, ...) they read
\ba
  R_V^a(t) &=& \int_{-1}^{1}\, \frac{d\bar{x}}{\bar{x}}\, 
                           H^a(\bar{x},0;t)\,,\nn\\
  R_A^a(t) &=& \int_{-1}^{1}\, \frac{d\bar{x}}{\bar{x}}\,{\rm sign}(\bar{x})\, 
                   \widetilde{H}^a(\bar{x},0;t)\,,\nn\\
  R_T^a(t) &=& \int_{-1}^{1}\, \frac{d\bar{x}}{\bar{x}}\, E^a(\bar{x},0;t)\,,
\label{softff}
\ea
where $\bar{x}=(k_j+k_j')^+/(p+p')^+$. The full form factors in
(\ref{final}), specific to Compton scattering, are given by
\be
             R_i(t) = \sum_a e_a^2\, R_i^a(t)\,,
\ee
$e_a$ being the charge of quark $a$ in units of the positron
charge. In principle there is a fourth form factor being related to 
the GPD $\widetilde{E}^a$ but it does not contribute to the Compton
amplitudes in the symmetric frame. The form factors $R_i^a$ also
appear in wide-angle photo- and electroproduction of mesons \cite{hanwen}.

\begin{figure}
\begin{center}
\epsfig{figure=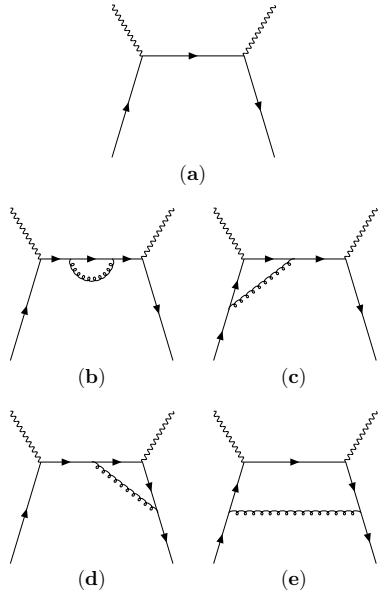, bb= 215 405 400 700, width=5.0cm}
\end{center}
\vspace{-0.4cm}
\caption{\label{fig2} Feynman graphs for Compton scattering off
on-shell quarks. a) is the LO graph, the others represent the NLO QCD
corrections. Graphs with self energy corrections to external fermions
and those with interchanged interaction points of the
photons are not shown.} 
\end{figure}

Last not least, the ${\cal H}_{\mu'\lambda',\,\mu\lambda}$  in Eq.\ 
(\ref{final}) denote the $\gamma q \to \gamma q$ subprocess amplitudes
where the helicities $\lambda$ and $\lambda'$ refer to the quarks now. 
To leading order (LO) these amplitudes are to be calculated from the
Feynman graphs shown in Fig.\ \ref{fig2} a. One finds
\be
{\cal H}^{LO}_{++,++}=2\sqrt{\frac{s}{-u}}\,, \qquad
{\cal H}^{LO}_{-+,-+}= 2\sqrt{\frac{-u}{s}}\,, \qquad
{\cal H}^{LO}_{-+,++}=0 \,.
\label{pqcd0}
\ee
Since the quarks are taken as massless there is no quark helicity
flip, ${\cal H}_{\mu'\lambda,\,\mu-\lambda}=0$ to any order of
$\als$. Other helicity amplitudes are obtained from those given in
(\ref{pqcd0}) by parity and time reversal invariance
\be
{\cal H}_{-\mu'-\lambda',-\mu-\lambda} \,=\,
{\cal H}_{\mu\lambda,\,\mu'\lambda'} \,=\, 
(-1)^{\mu-\lambda-\mu'+\lambda'}\, {\cal H}_{\mu'\lambda',\,\mu\lambda}\,.
\ee
Analogous relations hold for the ${\cal M}_{\mu'\nu',\,\mu\nu}$.

The NLO corrections to the $\gamma q\to \gamma q$ are to be calculated
from the Feynman graphs b - e depicted in Fig.\ \ref{fig2}. We work in
Feynman gauge and use dimensional regularization ($n=4 + \epsilon$). 
As expected for the process at hand, the ultraviolet divergencies
of the individual graphs cancel in the sum, the NLO amplitudes are ultraviolet
safe. On the other hand, those photon helicity non-flip amplitudes which
are non-zero at LO, are infrared (IR) divergent. They can be
decomposed into an infrared divergent part $\propto {\cal H}^{LO}$ and an infrared
safe one, ${\cal H}^{NLO}$,
\be
 {\cal H}_{\mu +,\, \mu +}^{IR} = \frac{\als}{4\pi}\, C_F\,
                     C_{IR}(\mu_F)\, {\cal H}_{\mu +,\, \mu +}^{LO}
                     + {\cal H}_{\mu +,\, \mu +}^{NLO}\,,
\label{decomp}
\ee
where $C_{IR}$ embodies the IR singularities.
$C_F$ ($=4/3$) is a colour factor and $\mu_F$ is a factorization scale 
being of order $\Lambda$. As usual, we interprete the infrared
divergent pieces as non-perturbative physics and absorb them into the
soft form factors. Thus, we write for any of the products of hard
scattering amplitudes and form factors appearing in (\ref{final})
\ba
 {\cal H}_{\mu +,\, \mu +}(s,t)\,R_i(t) &=& \left[{\cal H}_{\mu +,\, \mu +}^{LO}\,
                   \left(1 + \frac{\als}{4\pi}\, C_F\, C_{IR}(\mu_F)\right) 
                    +  {\cal H}_{\mu +,\, \mu +}^{NLO} \right]\, R_i(t) \nn\\
                   &=& \left[{\cal H}_{\mu +,\, \mu +}^{LO}\,
                        +  {\cal H}_{\mu +,\, \mu +}^{NLO} \right]\, R_i(t,\mu_F)
                        \,+\, {\cal O}(\als^2)\,.
\ea
The next issue we are concerned with is the exact definition of
$C_{IR}$. The infrared divergencies in (\ref{decomp}) have the form 
\be
  -\left(\frac{-t}{4\pi\mu_F^2}\right)^{\epsilon/2}\, \Gamma(1-\epsilon/2)\,
            (8/\epsilon^2 - 6/\epsilon)\,.
\label{infra}
\ee
The $1/\epsilon^2$ term appears as a consequence of overlapping soft
and collinear divergencies. The accompanying double logs become
large at large $-t$ and have to be resummed together with corresponding 
higher order terms $[\als \ln^2(-t/\mu_F^2)]^n$ in a Sudakov factor
\cite{collins}. The same problems occur in the Feynman
contribution to the electromagnetic form factor of the proton which is
the analogue of the handbag contribution to Compton scattering.
To NLO the $\gamma^*\to q \bar{q}$ vertex appearing in that calculation,
provides infrared singularities identically to (\ref{infra}) which
have to be absorbed into the soft hadronic matrix element, too. It is,
of course, natural to use the same scheme for the regularization of
the IR divergencies for both the Feynman and the handbag contribution.
Since customarily the Sudakov factor is considered as part of the 
electromagnetic form factor \cite{collins,rad}, i.e.\ the latter
already includes resummed double logs, we are forced to identify $C_{IR}$
with the full expression (\ref{infra}) in order to match with standard 
phenomenology and, in particular, with the model we employ in our
numerical studies of Compton scattering. We remark in passing that the 
$\gamma^*\to q \bar{q}$ vertex also occurs in $e^+ e^-\to q \bar{q}$. 
In this case the infrared singularities are compensated by real gluon 
emission. The infrared divergencies generated by the NLO QED corrections 
to Compton scattering off electrons cancel against those of double 
Compton scattering, $\gamma e\to \gamma\gamma e$ \cite{brown52}. 
In deeply virtual Compton scattering only a single IR pole appears
\cite{bel} but it can be shown that in the limit $x_j\to 1$ an
additional singularity emerges \cite{mue01}.

After removal of the IR divergencies the NLO amplitudes read:
\ba
{\cal H}_{++,++}^{NLO}&=& \frac{\als}{2\pi}C_F \left\{
                \frac{\pi^2}{3} -7 +
                 \frac{2t-s}{s}\ln\frac{t}{u}\right. \nn\\
           && \hspace{2cm} +\, \left. \ln^2\frac{-t}{s} + 
               \frac{t^2}{s^2}\left(\ln^2\frac{t}{u}
              +\pi^2\right)-2i\pi \ln\frac{-t}{s}\right\}
                                         \sqrt{\frac{s}{-u}}\,,\nn\\
{\cal H}_{-+,-+}^{NLO}&=& \frac{\als}{2\pi} C_F
              \left\{\frac43 \pi^2 - 7 
            +\frac{2t-u}{u}\ln\frac{-t}{s} + \ln^2\frac{t}{u}  
                                            \right.  \nn\\
         &&\left.\hspace{2cm} +\,\frac{t^2}{u^2}\ln^2\frac{-t}{s}-
                                     2i\pi\left(\frac{2t-u}{2u}
      +\frac{t^2}{u^2}\ln\frac{-t}{s}\right)\right\}\sqrt{\frac{-u}{s}}\,, \nn\\
{\cal H}_{-+,++}^{NLO}&=&-\frac{\als}{2\pi} C_F\left\{\sqrt{\frac{s}{-u}}
                     +\sqrt{\frac{-u}{s}}\right\}\,.
\label{pqcd1}
\ea
Since in wide-angle Compton scattering $-t$ and $-u$ are of order $s$
there are no large logs in the NLO amplitudes. We also see that the
NLO amplitudes possess both non-zero imaginary parts and non-zero
photon helicity flips. 

At the one-loop level, there is a complication which we have to
discuss next, namely gluons have to be considered as active partons as well. 
The treatment of the gluonic contributions to wide-angle Compton scattering
is analogous to that one utilized in wide-angle photo- and
electroproduction of vector mesons \cite{hanwen}. The gluonic
contributions factorize into the parton subprocess $\gamma g \to
\gamma g$ and gluonic form factors. In contrast to the case of quarks,
the partonic amplitudes now allow parton, i.e.\ gluon, helicity
flips. 

For gluon helicity non-flip the gluonic contributions have a
representation analogous to (\ref{final}). The corresponding 
form factors read
\be
 R_V^g (t) = \sum_a e^2_a \int_0^1 \frac{d\bar{x}}{\bar{x}^2}\, 
                                               H^g(\bar{x},0;t)\,,
\label{gluonff}
\ee
and analogously for the other ones. The range of integration is
restricted to the interval $[0,1]$ since gluons and antigluons are the
same particles. The additional factor $1/\bar{x}$ is conventional; it
appears as a consequence of the definition of the gluon GPDs
\cite{rad97,ji97,mue94} which implies the forward limits
\be
           \bar{x} g(\bar{x}) = H^g(\bar{x},0;0)\,, \qquad 
    \bar{x} \Delta g(\bar{x}) = \widetilde{H}^g(\bar{x},0;0)\,.
\label{gluon}
\ee
With regard to this definition we still term (\ref{gluonff}) a
$1/\bar{x}$-moment. The sum in (\ref{gluonff}) runs over the flavors 
$u$, $d$, $s$ which suffices for the range of energy we are interested
in. For gluon helicity flip we do not present details here because
these contributions are neglected in our numerical studies as those
proportional to the form factors $R_A^g$ and $R_T^g$. This is
justified since, as we will argue in Sec.\ \ref{sect3}, the
gluon form factors are expected to be smaller than the corresponding
quark ones at large $-t$ and since the gluonic
contributions only appear to order $\als$. Hence, we only consider the
contribution $\propto R_V^g$. It reads 
\be
            2\pi \aem \left[{\cal H}^g_{\mu'+,\,\mu +}(s,t) + {\cal
                       H}^g_{\mu'-,\,\mu -}(s,t)\right]\, R_V^g (t)\,, 
\ee
and is to be added to the proton helicity non-flip amplitudes 
${\cal M}_{\mu'+,\,\mu +}$ in (\ref{final}).  

\begin{figure}[t]
\begin{center}
\epsfig{figure=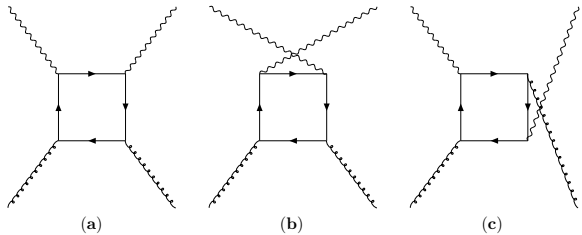, bb= 140 550 490 690, width=7.8cm}
\end{center}
\caption{\label{fig3} Sample Feynman graphs for photon-gluon scattering.} 
\end{figure}
The photon-gluon amplitudes are to be calculated from the three graphs
shown in Fig.\ \ref{fig3}. There are three further graphs contributing 
to order $\als$ which however reduce to first three ones by reversing the
fermion number flow. After some algebra we find for the gluon helicity 
non-flip amplitudes
\begin{eqnarray}
{\cal H}^g_{++,++}&=&\frac{\alpha_s}{\pi}\left\{\frac{t^2+u^2}{2s^2}
                     \left(\ln^2\frac{t}{u}+\pi^2\right)
                      +\frac{t-u}{s}\ln\frac{t}{u}+1\right\}\,, \nn\\
{\cal H}^g_{-+,-+}&=&\frac{\alpha_s}{\pi}\left\{\frac{s^2+t^2}{2u^2}\ln^2\frac{-t}{s}
                    +\frac{t-s}{u}\ln\frac{-t}{s}+1-
                    i\pi\left(\frac{t-s}{u}+\frac{s^2+t^2}{u^2} 
                    \ln\frac{-t}{s}\right)\right\}\,,\nn\\
{\cal H}^g_{-+,++}&=&-\frac{\alpha_s}{\pi}\,.
\label{pqcdg}
\end{eqnarray}
Except for a different normalization these amplitudes agree with
those given in Ref.\ \cite{bern}. In this recent paper the gluon
helicity flip amplitudes can be found, too.
\section{Modelling the GPDs}
\label{sect3}
In order to predict wide-angle Compton scattering a model for the
GPDs at large $-t$ and zero skewedness is required. In Ref.\
\cite{DFJK3} (see also \cite{link,bro01}) it has been shown on the 
basis of light-cone quantization that the GPDs possess a
representation in terms of light-cone wave function overlaps. This 
representation allows the construction of a simple model for the GPDs 
by parameterizing the transverse momentum dependence of a $N$-particle 
wave function as
\be
\Psi_N \propto \exp{\left[-a_N^2 \sum_{i=1}^N k_{\perp i}^2/x_i\right]} \,,
\ee
which is in line with the central assumption of the handbag approach
of restricted $k_{\perp i}^2/x_i$, necessary to achieve factorization
of the amplitudes into soft and hard parts. Without explicit
specification of the $x$-dependencies of the wave functions one can
then calculate the $\xi=0$ GPDs from the overlap representation if a 
common transverse size parameter $a=a_N$ is used. This ansatz leads to 
\be
H^a(\bar{x},0;t) = \exp{\left[a^2 t
        \frac{1-\bar{x}}{2\bar{x}}\right]}\, q_a(\bar{x})\,, \qquad 
\widetilde{H}^a(\bar{x},0;t) = \exp{\left[a^2 t
        \frac{1-\bar{x}}{2\bar{x}}\right]}\, \Delta q_a(\bar{x})\,,
\label{gpd}
\ee
where $q_a$ and $\Delta q_a$ are the ordinary unpolarized and polarized 
parton distributions for a quark of flavour $a$, respectively. An 
analogous representation holds for the gluon GPDs with the replacement 
of $q_a$ and  $\Delta q_a$ by $\bar{x} g$ and $\bar{x} \Delta g$, respectively.

\begin{figure}[t]
\begin{center}
\epsfig{file=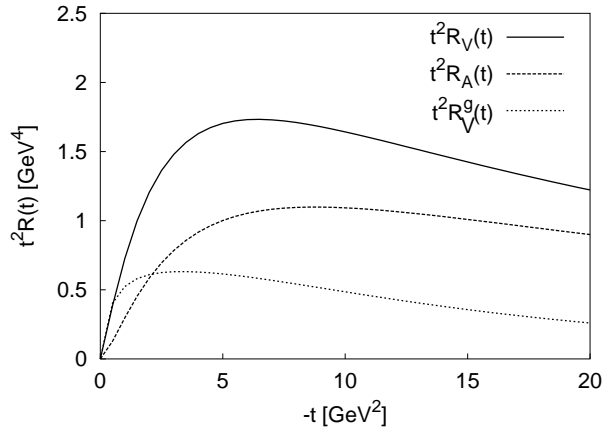, bb=50 45 397 295, width =8cm}
\end{center}
\vspace{-0.4cm}
\caption{\label{form} The Compton form factors $R_V$, $R_A$ and
$R_V^g$, scaled by $t^2$, vs.\ $-t$.}
\end{figure}

Taking the parton distributions from one of the current analyses of
deep inelastic lepton-nucleon scattering, e.g.\ from Ref.\ \cite{grv},
and using a value of $\simeq 1 \gev^{-1}$ for the transverse size
parameter $a$, one obtains acceptable results for the unpolarized
Compton cross section as well as for the proton and neutron
electromagnetic Dirac form factors, $F_1$, which represent $x^0$-moments 
of $H^a$. The model GPDs (\ref{gpd}) have been improved somewhat by
treating the lowest three proton Fock states explicitly with specified 
wave functions \cite{link,bolz} whose parameters are fitted to data
for the electromagnetic form factors and to the parton distributions
given in \cite{grv}. Due to this procedure the form factors
effectively include the Sudakov factors and do practically not 
depend on the factorization scale. Since we are merely interested in a
restricted range of momentum transfer we ignore the evolution of the
GPDs as has been done in previous work \cite{rad98,link,hanwen}. As 
shown by Vogt \cite{vogt}, the evolution can be incorporated in the 
overlap model for the GPDs at the expense of a scale dependent 
transverse size parameter. Numerical results for the form factors, 
obtained from the improved version of the overlap model \cite{link,hanwen}, 
are displayed in Fig.\ \ref{form}. We will employ these results
in our numerical studies. 

Let us now discuss the form factor $R_T$. The overlap representation
of the underlying GPD $E^a$ involves components of the proton wave 
functions where the parton helicities do not add up to the helicity 
of the proton. In other words, parton configurations with non-zero 
orbital angular momentum contribute to it. That $E^a$ involves parton 
orbital angular momentum in an essential way is also reflected in 
Ji's angular momentum sum rule \cite{ji98}. Whereas $R_V$ and $F_1$ 
represent different moments of the GPD $H^a$, correspond $R_T$ and 
the Pauli form factor, $F_2$, to $E^a$. Since, at large $-t$, 
the integrals (\ref{softff}) for $R_V$ and $R_T$ as well as
those for their electromagnetic counter parts, $F_1$ and $F_2$, are 
dominated by the region $\bar{x}\to 1$ where the valence
$u$-quarks provide most of the contributions, there is little 
difference between $1/x$ and $x^0$-moments. This is sufficiently
suggestive to assume that

\be
                    R_T/R_V \simeq F_2/F_1\,.
\ee   
Inspection of the SLAC data on $F_2$ \cite{and94} therefore 
leads one to the expectation $R_T/R_V \propto \Lambda^2/t$ with the
consequence of parameterically suppressed ($\propto\Lambda/\sqrt{-t}$)
contributions from $R_T$ to the Compton amplitudes (\ref{final}). Given 
that already the evaluation of the handbag diagram is only accurate up 
to corrections of order $\Lambda^2/t$, $R_T$ and consequently proton 
helicity flip is to be neglected for consistency. 
This has been done in previous LO calculations \cite{rad98,link}. 
However, the recent JLab measurement of $F_2$ \cite{jon00} seems to 
indicate a behaviour $\propto \Lambda/\sqrt{-t}$ for the ratio of form
factors rather than $\Lambda^2/t$. Provided this behaviour will be 
confirmed, $R_T$ cannot be omitted in the handbag approach; it 
contributes to the same order in $\Lambda/\sqrt{-t}$ as the other 
form factors, see (\ref{final}). The results for Compton scattering 
presented in \cite{rad98,link} have to be revised accordingly.
Note that a behaviour $\propto \Lambda/\sqrt{-t}$ for the ratio of
form factors appears quite natural in the overlap representation 
\cite{DFJK3,kroll01}. 

In the next section we will present predictions for Compton scattering
using both the scenarios $R_T$ omitted and $R_T/R_V \propto 
\Lambda/\sqrt{-t}$ for comparison. In the latter case we use
a value of 0.37 for the ratio
\be
        \kappa = \frac{\sqrt{-t}}{2m}\, \frac{R_T}{R_V}\,,
\ee 
as taken from the experimental ratio of $F_2$ and $F_1$
measured by the JLab Hall A Collaboration \cite{jon00}. 
In general $\kappa$ is a function of $t$.

The gluonic form factors play a minor role in our analysis since
they contribute only to order $\als$. Moreover, they are smaller than
their quark counterparts at large $-t$ since there, as we argued
above, the form factors are controlled by the region $\bar{x}\simeq 1$
where the valence $u$-quark dominates. $R_T^g$, related to $E^g$, as 
well as the gluon helicity flip form factors \cite{diehl01} involve 
parton orbital angular momentum. One may therefore anticipate that 
these form factors are smaller than $R_V^g$. $R_A^g$, being related 
to $\Delta g$, is expected to be very small, too \cite{hanwen}. Thus, 
only the largest of the gluonic form factors, $R_V^g$, is taken into 
account by us, the other ones are neglected. Numerical results for 
$R_V^g$ are taken from \cite{hanwen} and shown in Fig.\ \ref{form}.
\section{Observables for Compton scattering off protons}
\label{sect4}
The derivation of the Compton amplitudes within the handbag approach
naturally requires the use of the light-cone helicity basis. However,
for comparison with experimental and other theoretical results the use
of the ordinary photon-proton c.m.s.\ helicity basis is more
convenient. The c.m.s.\ helicity amplitudes $\Phi_{\mu'\nu',\,\mu\nu}$
(we keep the notation of the helicity labels) are obtained from the
light-cone helicity amplitudes (\ref{final}), defined
in the symmetric frame, by the following transform \cite{diehl01}
\be 
\Phi_{\mu'\nu',\,\mu\nu} \,=\, {\cal M}_{\mu'\nu',\,\mu\nu}
               +\beta/2 \left[\, (-1)^{1/2-\nu'} {\cal M}_{\mu'-\nu',\,\mu\nu}
                  +  (-1)^{1/2+\nu} {\cal M}_{\mu'\nu',\,\mu -\nu}\,\right]
                  + {\cal O}(\Lambda^2/t) \,,
\label{trans}
\ee
where 
\be
              \beta = \frac{2m}{\sqrt{s}}\, \frac{\sqrt{-t}}{\sqrt{s}
                           + \sqrt{-u}} \,.
\ee
For convenience we will use below a more generic notation for the six
independent helicity amplitudes \cite{rollnik}
\ba
\Phi_1 &=& \Phi_{++++}\,, \qquad \Phi_3 = \Phi_{-+++}\,, \qquad 
                               \Phi_5 = \Phi_{+-+-}\,, \nn\\  
\Phi_2 &=& \Phi_{--++}\,, \qquad \Phi_4 = \Phi_{+-++}\,, 
                                         \qquad \Phi_6 = \Phi_{-++-}\,.
\ea 
Inspection of (\ref{trans}) and (\ref{final}) reveals that 
\be
              \Phi_2 = - \Phi_6 \,+\, {\cal O}(\Lambda^2/t)\,,
\label{phi26}
\ee
within the handbag approach. The amplitudes $\Phi_2$, $\Phi_3$ and
$\Phi_6$ are of order $\als$.

In our numerical studies we choose $s/2$ as the scale of $\als$ which
is the typical virtuality one encounters in the Feynman graphs shown
in Figs.\ \ref{fig2} and \ref{fig3}, and
evaluate $\als$ from the two-loop expression for $n_f=4$ flavours and
$\Lambda_{\overline{MS}}^{(4)}=305 \mev$ \cite{PDG}. We emphasize that
our predictions, termed scenario A in the following, include
corrections of order $\als$ and $\beta$ ($\propto\Lambda/\sqrt{-t}$)
as well as contributions from $R_T$ (with $\kappa=0.37$). Terms of
order $\als^2$ and  $\beta^2$ are neglected throughout. Thus, for
instance, a square of a helicity amplitude is evaluated as 
\be
         |{\cal H}|^2= |{\cal H}^{LO}|^2 + 
                       2\, {\cal H}^{LO}\, {\rm Re}\,{\cal H}^{NLO}\,.
\ee
For comparison we also show the results given in \cite{link}
where only the LO subprocess amplitudes are taken into account and
$R_T$ as well as the order $\Lambda/\sqrt{-t}$
corrections are omitted (scenario B).

\begin{figure}[bt]
\begin{center}
\epsfig{file=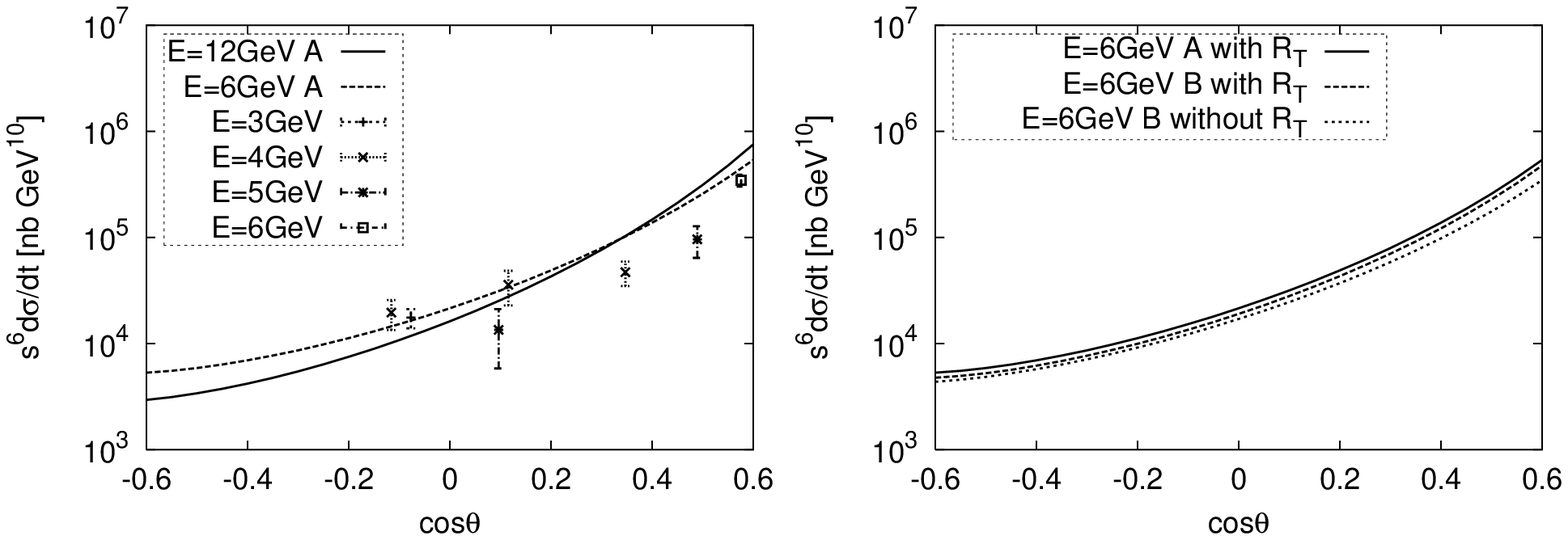, bb= 45 55 320 250, width=10cm, clip=}
\end{center}
\vspace{-0.4cm}
\caption{\label{cross} Predictions from scenario A for the Compton 
cross section, scaled by $s^6$, versus $\cos\theta$ for various photon 
energies, $E$, in the laboratory frame. $\theta$ is the c.m.s.\ scattering 
angle. Data taken from Ref.\ \cite{shupe}; they are only shown for 
$-t,-u\ge 2.5\; \gev^2$ as a minimum condition for our approximations to 
be applicable.} 
\end{figure}
\begin{figure}[tb]
\begin{center}
\epsfig{file=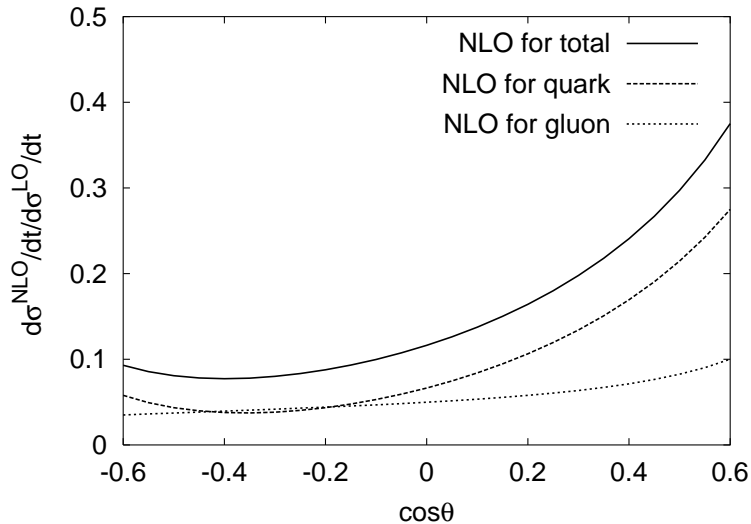, bb= 50 50 400 295, width=10cm, clip=}
\end{center}
\vspace{-0.4cm}
\caption{\label{crossd} Comparison between the NLO and LO results for
the Compton cross section (contribution from $R_T$ not included) 
at a photon lab. energy of 6 GeV.} 
\end{figure}

The simplest but most important observable is the unpolarized cross
section. In terms of the c.m.s.\ helicity amplitudes and within the
handbag approach it reads:
\ba
\frac{d\sigma}{dt}&=& \frac{1}{32\pi(s-m^2)^2} \left[ |\Phi_1|^2 +
        |\Phi_2|^2 + 2|\Phi_3|^2 + 2|\Phi_4|^2 + |\Phi_5|^2 +
                                         |\Phi_6|^2 \right] \nn\\
  &=& \frac{\pi\aem^2}{4(s-m^2)^2} \bigg\{  R_V^2\, \big[ 1 + \kappa^2 \big] 
         \big|{\cal H}_{++++} + {\cal H}_{+-+-} \big|^2  
            + R_A^2\, \big|{\cal H}_{++++} - {\cal H}_{+-+-}
                      \big|^2  \nn\\
    && \hspace{2cm} + 2\, ({\cal H}_{++++}^{LO} 
            + {\cal H}_{+-+-}^{LO})\, {\rm Re}\, ({\cal H}_{++++}^g  
              + {\cal H}_{+-+-}^g)\, R_V R_V^g \bigg\} \,,
\label{eq:cross}
\ea
where we keep the proton mass in the phase space factor. In Fig.\
\ref{cross} we compare our scenario A results for the Compton cross
section, scaled by $s^6$, with experiment \cite{shupe}. This scaling 
accounts for most of the energy dependence in the kinematical range 
of interest. 
As can be seen from Fig.\ \ref{form}, the form factors behave as
$1/t^2$ in the momentum transfer range from about 5 to 15 GeV$^2$ and, 
consequently, the Compton cross section exhibits approximate
dimensional counting rule behaviour ($\propto s^{-6}$) at fixed 
scattering angle in a limited range of energy. With increasing $-t$
the form factors gradually turn into a $\propto t^{-4}$ behaviour. 
In that region of $t$, likely well above $100\, \gev^2$ as is argued 
in \cite{link}, the perturbative contribution to Compton scattering 
will take the lead.  
For our form factor model the contribution from $R_T$
results in a constant factor of 1.13 multiplying $R_V^2$ while that
from $R_A$ is very small in the forward hemisphere and grows to about 
$17\%$ for $\cos\theta\simeq -0.6$. This comes about because 
$|{\cal H}_{++++}+{\cal H}_{+-+-}|^2 \gg |{\cal H}_{++++}-{\cal H}_{+-+-}|^2$
in the wide angle region and because, according to the overlap model,
$R_V > R_A$. 
In order to demonstrate the importance of the NLO corrections we 
display ratios of the NLO corrections for quarks and gluons and the 
LO result in Fig.\ \ref{crossd}. As can be seen from that figure 
the gluonic contribution amounts to less than $10\%$ in the entire 
$\cos\theta$ range of interest. The NLO corrections from quarks 
are small in the backward hemisphere while the grow up to about $30\%$
for $\cos\theta\to 1$. This happens because $s$ and 
$-t$ differ greatly for $\cos \theta \to 0.6$ and, hence, some of the 
logs in (\ref{pqcd1}) become large. $\cos \theta \simeq 0.6$ is the border
line for the applicability of the handbag approach beyond which $-t$
can hardly be regarded as being of order $s$. 

Given the quality of the data, and the small
energies and low values of $-t$ and $-u$ at which they are available, 
the predictions following from the handbag approach are in fair 
agreement with experiment. Better data are clearly needed for a
severe test of the handbag approach and its confrontation with other 
approaches. Cross sections of comparable magnitude have been
obtained within a diquark model \cite{kro91}. This model is a
variant of the perturbative approach in which diquarks are considered as
quasi-elementary constituents of the proton \cite{diquarks}.
In the leading-twist perturbative approach, on the other hand, it
seems difficult to account for the Compton data even if strongly  
asymmetric distribution amplitudes are used \cite{dixon}. For more 
symmetric ones, like the asymptotic one or that one proposed in
\cite{bolz} the perturbative predictions are way below
experiment \cite{dixon}.
 
Before we turn to the discussion of spin dependent observables a remark
concerning the definition of the proton polarization states is in
order. We use the convention advocated by Bourrely, Leader and Soffer 
\cite{BLS} and define the rotation of a vector through an azimuthal angle
$\varphi$ and a polar angle $\theta$ by the matrix $R(\varphi,
\theta, 0)$. We consider three different polarization states of the
proton -- $L$, $S$ and $N$ -- defined as spin eigenstates of ${\bf
A}\cdot {\bf \sigma}$ where ${\bf \sigma}$ is the vector formed of the
Pauli matrices and ${\bf A}$ any of the unit vectors
\be
{\bf L}^({}'^) = \frac{{\bf p}^({}'^)}{|{\bf p}^({}'^)|}\,, \qquad 
{\bf N} = {\bf L} \times {\bf L}'\,, \qquad
{\bf S}^({}'^) = {\bf N} \times {\bf L}^({}'^)\,.
\label{vectors}
\ee
${\bf p}$ and ${\bf p}'$ denote the three-momenta of the incoming and
outgoing protons, respectively. For Compton scattering a number of 
polarization observables have been introduced in order to probe  
theoretical ideas \cite{rollnik}, many more can be defined in principle. 
Obviously, only a few of them can be discussed here. 

\begin{figure}[t]
\begin{center}
\psfig{file=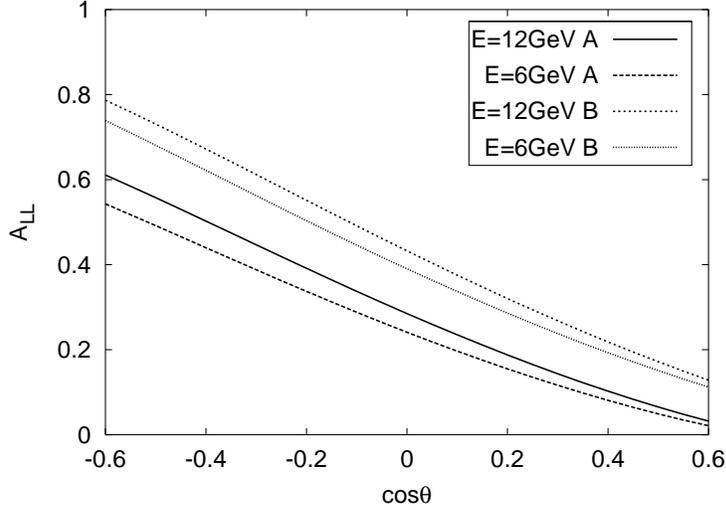, width =10cm}
\end{center}
\vspace{-0.4cm}
\caption{\label{all} Predictions for the initial state helicity
  correlation $A_{\rm LL}$ from scenario A and B at photon
  lab.\ energies of 6~\gev{} and 12~\gev{}.}
\end{figure}

One set of polarization observables are the two-spin correlations of
which the helicity ($L$-type) correlations are of particular
interest. That one of the photon and the proton in the 
initial state is defined by
\ba
A_{LL}\frac{d\sigma}{dt}
      &=&\frac{1}{2}\left[\frac{d\sigma(++)}{dt}-\frac{d\sigma(+-)}{dt}\right]
                                              \nn\\
      &=& \frac{1}{32\pi(s-m^2)^2} \left[ |\Phi_1|^2 +  |\Phi_2|^2
                       - |\Phi_5|^2 - |\Phi_6|^2 \right] \nn\\
      &=& \frac{\pi\aem^2}{2(s-m^2)^2} R_A\, \bigg\{
             R_V\, \Big[1 -\beta \kappa \Big]\,
            \Big[\big|{\cal H}_{++++}\big|^2 -\big| {\cal H}_{+-+-}\big|^2
                         \Big]  \nn\\
      && \hspace{0.5cm} +\; R_V^g\, \big({\cal H}_{++++}^{LO} 
            - {\cal H}_{+-+-}^{LO}\big)\, {\rm Re}\,\big({\cal H}_{++++}^g  
              + {\cal H}_{+-+-}^g\big)\, \bigg\} \,.
\ea
Using the model form factors discussed in Sect.\ \ref{sect3}, we
evaluate the initial state helicity correlation $A_{LL}$ for scenario
A and compare it in Fig.\ \ref{all} to that obtained from scenario B
\cite{link}. The $\cos\theta$ dependence of $A_{LL}$ approximately
reflects that of the corresponding helicity correlation for the 
photon-parton subprocess, $(s^2-u^2)/(s^2+u^2)$, its size being however
diluted by the form factors. We observe from Fig.\ \ref{all} that the 
inclusion of $R_T$ and the NLO corrections reduce the values of
$A_{LL}$ by about $0.1$ to $0.2$ as compared to the results from
scenario B. The results from the handbag approach are opposite in sign
to the diquark model predictions \cite{kro91}. In the leading-twist
perturbative approach, the results for $A_{LL}$ are also markedly
different from our ones \cite{dixon}. They are very sensitive to the
proton distribution amplitudes used in the evaluation. 

The analogous correlation between the helicities of the incoming
photon and the outgoing proton is defined by
\ba
K_{LL}\frac{d\sigma}{dt}
      &=& \frac{d\sigma(++)}{dt}-\frac{d\sigma(+-)}{dt} \nn\\
      &=& \frac{1}{32\pi(s-m^2)^2} \left[ |\Phi_1|^2 -  |\Phi_2|^2
                       - |\Phi_5|^2 + |\Phi_6|^2 \right]\,. 
\ea
Since $\Phi_2=-\Phi_6$ in the handbag approach, see (\ref{phi26}), we
obtain 
\be
             K_{LL}=A_{LL}\,. 
\ee
The helicity transfer from the
incoming to the outgoing photon reads 
\ba
D_{LL}\frac{d\sigma}{dt}&=& \frac{d\sigma(++)}{dt}-\frac{d\sigma(+-)}{dt}\nn\\
                        &=&\frac{1}{32\pi(s-m^2)^2} \left[ |\Phi_1|^2 
            - |\Phi_2|^2 - 2|\Phi_3|^2 + 2|\Phi_4|^2 + |\Phi_5|^2 
                                        - |\Phi_6|^2 \right] \,.
\label{eq:dll}
\ea                                       
Although the photon helicity is not strictly conserved to NLO,
the helicity transfer is
\be 
                            D_{LL} = 1 + {\cal O} (\als^2)\,,
\ee 
as is evident from a comparison of (\ref{eq:cross}) and (\ref{eq:dll}).

\begin{figure}[t]
\begin{center}
\psfig{file=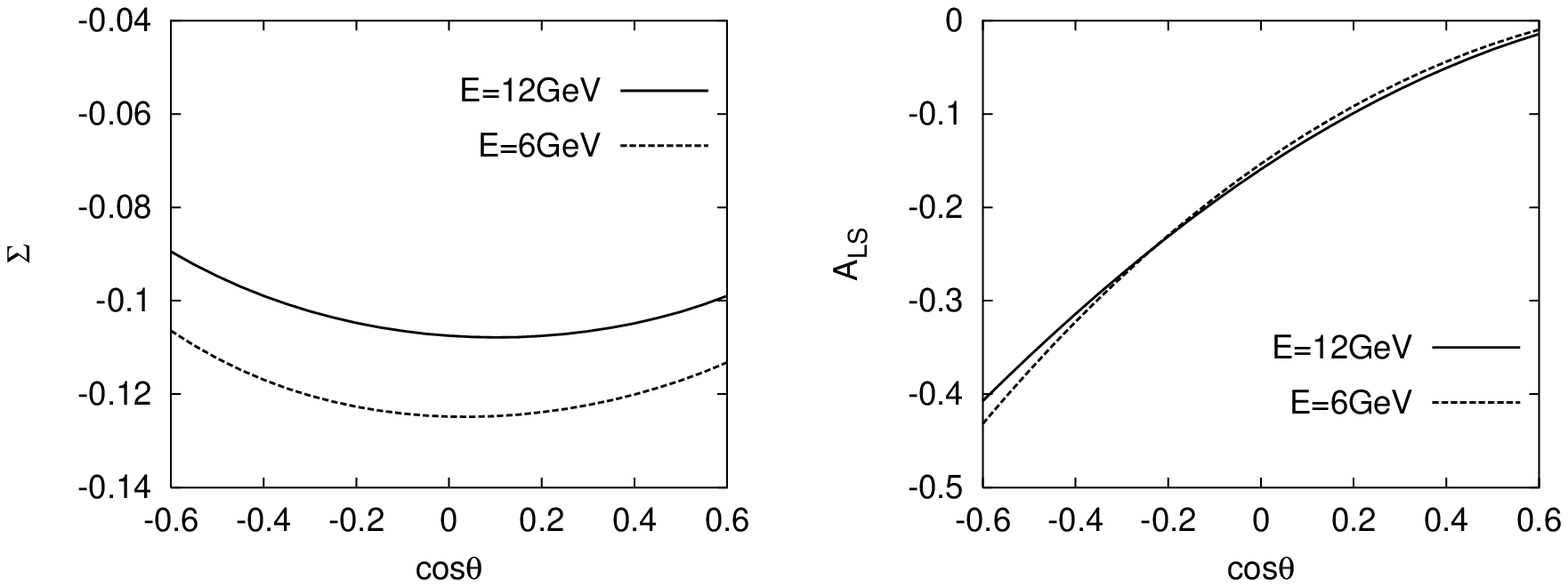, bb=323 55 565 245, width =9cm, clip=}
\end{center}
\vspace{-0.4cm}
\caption{\label{als} The correlation $A_{LS}$  
at photon lab.\ energies of 6 GeV and 12 GeV.}
\end{figure}

One may also consider sideway proton spin directions, see (\ref{vectors}).
The correlation between the helicity of the incoming photon
and the sideway ($S$-type) polarization of the incoming proton, parallel
($\rightarrow$) or antiparallel ($\leftarrow$) to the $S$-direction reads
\ba\label{eq:als}
A_{LS} \frac{d\sigma}{dt} &=& \frac12 \left[ \frac{d\sigma(+\rightarrow)}{dt}
                       - \frac{d\sigma(-\rightarrow)}{dt} \right] \nn\\
     &=& \frac{1}{16\pi (s-m^2)^2}\,{\rm Re}\,\left[\,(\Phi_1 -\Phi_5) \Phi_4^*
                             - (\Phi_2 +\Phi_6) \Phi_3^*\,\right] \nn\\
     &=& -\frac{\pi\aem^2}{2(s-m^2)^2}\,
             R_A\, \bigg\{\frac{\sqrt{-t}}{2m}\,R_T\,\Big[1 + \beta \kappa^{-1}\Big]\,
           \Big[\big|{\cal H}_{++++}\big|^2 - \big|{\cal H}_{+-+-}\big|^2
                                  \Big]       \nn\\
      && + \quad \beta\, R_V^g\, \big({\cal H}_{++++}^{LO} - 
                                       {\cal H}_{+-+-}^{LO}\big)\,
          {\rm Re}\, \big({\cal H}_{++++}^{g} + {\cal H}_{+-+-}^{g}\big)
                                   \bigg\} \,.
\ea
Predictions from scenario A are shown in Fig.\ \ref{als}, those
obtained from scenario B are zero. $A_{LS}$ turns out to
be rather independent of the photon energies. It is important to note
that $A_{LS}$ is a observable that is very sensitive to the form factor
$R_T$. The corrections from the term $\beta \kappa^{-1}$ are, however,
substantial, in particular for the energies available at JLab; they
cannot be ignored. This, after all, is the reason why, in
contrast to \cite{rad98,link}, we keep these terms. Neither in the 
diquark model \cite{kro91} nor in the leading-twist perturbative 
approach \cite{dixon} this observable has been discussed. 

The correlation between the helicity of the incoming photon and the 
sideway polarization of the outgoing proton is defined as
\ba
K_{LS} \frac{d\sigma}{dt} &=&  \frac{d\sigma(+\rightarrow)}{dt}
                       - \frac{d\sigma(-\rightarrow)}{dt}  \nn\\ 
     &=& -\frac{1}{16\pi (s-m^2)^2}\,{\rm Re}\,\left[\,(\Phi_1 -\Phi_5) \Phi_4^*
                             + (\Phi_2 +\Phi_6) \Phi_3^*\,\right]\,. 
\ea
Because of $\Phi_2=-\Phi_6$ it follows that
\be 
       K_{LS} = - A_{LS}\,. 
\ee
Correlations between the helicity of the incoming photon and the 
transverse ($N$-type)  polarization of either the incoming or the 
outgoing proton are zero due to parity invariance
\be
             K_{LN} = A_{LN} =0\,.
\ee 

\begin{figure}[t]
\begin{center}
\psfig{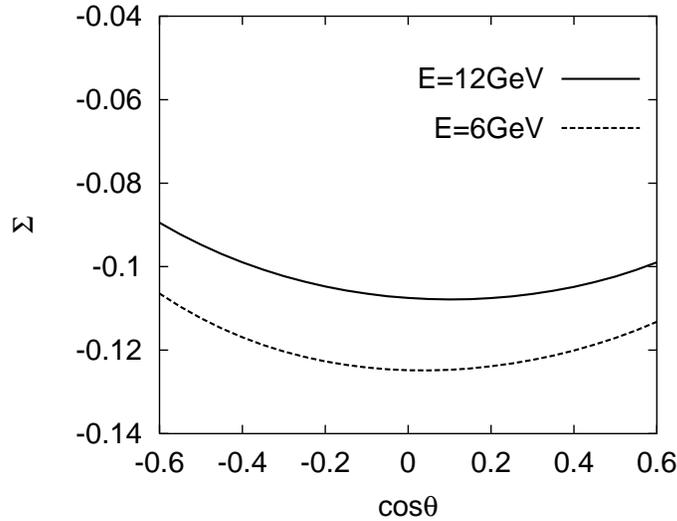}
\end{center}
\vspace{-0.4cm}
\caption{\label{asy} The incoming photon asymmetry $\Sigma$  
at photon lab.\ energies of 6 GeV and 12 GeV.} 
\end{figure}

A single-spin observable for Compton scattering is the incoming photon
asymmetry $\Sigma$ which is defined as
\ba \label{sum}
\Sigma\frac{d\sigma}{dt}&=&\frac{1}{2}\left[\frac{d\sigma_\perp}{dt}
                           -\frac{d\sigma_\parallel}{dt}\right] \nn\\
            &=&\frac{1}{16\pi (s-m^2)^2}\,{\rm Re}\, 
              \left[ \, (\Phi_1+\Phi_5) \Phi_3^*
                          + (\Phi_2-\Phi_6) \Phi_4^* \, \right]\nn\\
       &=& \frac{\als\,\aem^2}{(s-m^2)^2}\,  \frac{(s-u)^2}{us}\, 
  \Big[C_F\, R_V^2\, (1+\kappa^2) + 2\frac{\sqrt{-us}}{s-u} R_V R_V^g \Big]\,,
\ea
where $\perp$ and $\parallel$ refer to linear photon polarization
normal to and in the scattering plane, respectively. Obviously, 
$\Sigma$ is zero to LO since there is no photon helicity flip. The 
predictions obtained from scenario A are shown in Fig.\ \ref{asy}. 
$\Sigma$ is negative and small in absolute value. Approximately, 
i.e.\ if the terms $\propto R_A$ and $\propto R_V^g$ are neglected 
in (\ref{eq:cross}), it is given by
\be
 \Sigma \simeq - \als C_F/ \pi \,.
\ee 
Hence, the incoming photon asymmetry is nearly independent of the
Compton form factors. In the diquark model \cite{kro91} 
$\Sigma$ is negative too but smaller in absolute value. The leading-twist
approach \cite{dixon}, on the other hand, provides rather large 
positive values for $\Sigma$.

Last not least we want to comment on the ($N$-type) polarization or,
as occasionally termed, the single spin asymmetry of
the incoming proton, that of the outgoing one is analogous. 
The polarization of the incoming proton is defined by
\ba
P\frac{d\sigma}{dt} &=& \frac12 \left[ \frac{d\sigma(\uparrow)}{dt}
                          - \frac{d\sigma(\downarrow)}{dt} \right] \nn\\
        &=& \frac{1}{16\pi (s-m^2)^2}\,{\rm Im}\,\left[\,(\Phi_1+\Phi_5) \Phi_4^*
                             - (\Phi_2-\Phi_6) \Phi_3^* \,\right]\,,
\ea
where $\uparrow$ and $\downarrow$ denote the proton polarization
parallel and antiparallel to the $N$-direction, respectively.
The calculation of that polarization is a notoriously difficult task within
QCD. Therefore, many experimentally observed polarization effects, as
for instance the polarization in proton-proton elastic scattering at
large momenta transfer \cite{krisch}, remained unexplained. As is
well-known a non-zero polarization requires proton helicity flip 
and phase differences between the various helicity amplitudes. 
Both the necessary ingredients are provided by the handbag approach,  
helicity flips from $R_T$ and phases from the NLO corrections and we
approximately obtain
\be
  P \simeq - \frac{\sqrt{-t}}{2m}\, \frac{R_T\, R_V^g}{R_V^2}\,
                      \frac{\sqrt{-us}}{s-u}\, 
          {\rm Im}\, \big ({\cal H}_{++++}^g + {\cal H}_{+-+-}^g \big)\,.
\ee
The polarization is of order $\als$ and proportional to the gluonic 
contribution. Numerically it is very small, less than $3\%$ for our 
model form factors. The predictions for $P$ are to be taken with a
grain of salt. The neglect of gluon helicity flip as well as $\als^2$ 
and $\Lambda^2/t$ terms may lead to substantial corrections. Thus, at 
a conservative estimate, we can only say that an experimentally observed 
polarization larger in absolute value than, say, 0.1 - 0.2 near 
$\theta=90^\circ$ would be difficult to understand in the handbag approach.
\section{Measuring the Compton form factors}
In the preceding sections we presented predicitions for various
observables of wide-angle Compton scattering within the handbag
approach, using a model for the form factors that is based on
light-cone wave function overlaps \cite{link,DFJK3}. On the other 
hand, a model-independent test of the handbag approach is provided 
by a measurement of the Compton form factors which can be performed 
through an analysis of the data for a set of observables to be at 
disposal for several values of $s$ and $t$. The crucial question 
is whether or not the experimentally determined form factors are 
independent of $s$ within the experimental and theoretical 
uncertainties. 

At JLab the E99-114 collaboration plans to measure along with the
differential cross section the two-spin correlations $K_{LL}$ and
$K_{LS}$ \cite{nathan}. Provided the quality of this data will be 
sufficiently good one may isolate the three form factors $R_V(t)$, 
$R_A(t)$ and $R_T(t)$ (or $\kappa(t)$) from it. As a first step 
towards a model-independent analysis, one may neglect the gluonic 
contributions everywhere and the term $\propto R_A$ in the cross
section which, as we discussed above, is small.
To the extend that these simplifications are justified, one finds 
\ba
    \frac{d\sigma}{dt} &\simeq& \frac{\pi \aem^2}{4(s-m^2)^2} \,
        R_V^2(t) [1 + \kappa^2(t)]\, 
                 \big| {\cal H}_{++++} + {\cal H}_{+-+-}\big|^2\,,\nn\\
       K_{LL} &\simeq& 2 \, \frac{R_A(t)}{R_V(t)} 
                  \, \frac{1 -\beta \kappa(t)}{1 + \kappa^2(t)} \,
     \frac{\big|{\cal H}_{++++}\big|^2 - \big|{\cal H}_{+-+-}\big|^2}
        {\big|\; {\cal H}_{++++} + {\cal H}_{+-+-}\;\big|^2} \,,\nn\\
       \frac{K_{LS}}{K_{LL}} &\simeq& \kappa(t) \, 
         \frac{1 + \beta \kappa^{-1}(t)}{1-\beta \kappa(t)}\,.
\label{eq:appr}
\ea
The cross section is essentially controlled by the form factor $R_V$
with, probably, only a small correction from $\kappa$. $K_{LL}$ 
measures the ratio $R_A/R_V$ with, however, substantial corrections 
from $\kappa$. The ratio $K_{LS}/K_{LL}$ determines the Compton analogue 
$R_T/R_V$ ($\propto \kappa$) to the ratio of the electromagnetic form 
factors $F_2$ and $F_1$. For large energies and scattering angles near  
$90^\circ$, the $\beta$ terms are negligible small and the analysis is
markedly simplified. In Tab.\ \ref{tab1} we present an assessment of the
quality of the approximations (\ref{eq:appr}). The discrepancies
between (\ref{eq:appr}) and the full results from scenario A do not
exceed $15 \%$ at a photon energy of 6 GeV. The use of the LO
amplitudes in (\ref{eq:appr}) instead of the full ones enlarges the 
discrepancies, in particular in the forward hemisphere, see 
Fig.\ \ref{crossd}. The form factors measured through (\ref{eq:appr}) 
may be improved iteratively.

\begin{table}
\begin{center}
\begin{tabular}{|c|c|c|c|}  \hline
\phantom{\Big[}$\cos \theta$ & $\Delta (d\sigma/dt)$ [$\%$] & $\Delta (K_{LL})$
[$\%$] & $\Delta (K_{LS}/K_{LL})$ [$\%$] \\ \hline
\phantom{-} 0.6 & \phantom{1}-6.2 & \phantom{-}15.0 & \phantom{1}1.4 \\
\phantom{-} 0   & \phantom{1}-5.6 & \phantom{-1}2.4 & \phantom{1}2.2 \\
-0.6            & -14.8           & -11.2           &      14.6 \\
\hline
\end{tabular}
\end{center}
\caption{\label{tab1} The discrepancies between the approximations
(\ref{eq:appr}) and the full results from scenario A in percent of the
full results at a photon lab. energy of 6 GeV.}
\end{table}
\section{Summary}
As a complement to \cite{link} we have calculated the NLO QCD
corrections to the subprocess amplitudes and include the form factor
$R_T$, related to the GPD ${E}$, in the analysis of wide-angle 
Compton scattering off protons. We have also considered the
difference between the light-cone helicity basis in which the handbag
graph is calculated, and the usual c.m.s.\ helicity one. Predictions 
for various Compton observables are given and compared to the leading 
contribution discussed in \cite{link}. It turns out that these 
corrections are non-negligible in general although not unreasonably 
large. The NLO corrections and those due to the change of the helicity 
basis decrease with increasing energy while those due to  the form 
factor $R_T$ keep their size provided $\kappa$ is independent of $t$. 
We stress that there are uncontrolled corrections of order 
$\Lambda^2/t$ in the handbag approach. For energies as low as, say, 
3 GeV these corrections may be substantial. Our study may be of 
importance for severe tests of the handbag approach with future 
high-quality data for wide-angle Compton scattering which might be 
obtained at Spring-8, JLab or an ELFE-type accelerator.

\section*{Acknowledgments}
One of the authors (H.W.H.) would like to thank the Monbusho's Grand-in-Aid 
for the JSPS postdoctoral fellow for financial support. P.K. wishes to
acknowledge discussions with Markus Diehl, Rainer Jakob and Dieter M\" uller.

\end{document}